\documentclass[11pt,a4paper]{article}
\pdfoutput=1
\usepackage{jcappub}
\usepackage{subfigure}
\newcommand{\eq}[1]{\begin{equation}#1\end{equation}}
\newcommand{\eqa}[1]{\begin{eqnarray}#1\end{eqnarray}}
\newcommand{\secs}[1]{\section{#1\label{sec-#1}}}
\newcommand{\ssecs}[1]{\subsection{#1\label{ssec-#1}}}

\newcommand{\fig}[4]{\begin{figure}[#4]\centering\includegraphics[width=#3\textwidth]{Graph-#1.pdf}\caption{#2}\label{fig-#1}\end{figure}}
\newcommand{\figa}[3]{\begin{figure}[#3]\centering #1\caption{#2}\end{figure}}
\newcommand{\figi}[3]{\subfigure[#2]{\includegraphics[width=#3\textwidth]{Graph-#1.pdf}\label{fig-#1}}}
\newcommand{\tab}[4]{\begin{table}[#2]\centering\caption{#1\label{tab-#1}}\vspace{0.1in}\begin{tabular}{#3}#4\end{tabular}\end{table}}
\newcommand{\refeq}[1]{Eq.\ (\ref{eq-#1})}
\newcommand{\refsec}[1]{Section \ref{sec-#1}}

\newcommand{\refig}[1]{Fig \ref{fig-#1}}
\newcommand{\reftab}[1]{Table\ \ref{tab-#1}}

\newcommand{\subs}[1]{_\mathrm{#1}}
\newcommand{\sups}[1]{^\mathrm{#1}}
\newcommand{\dd}[1]{d#1}
\newcommand{\mpl}{M\subs{p}}

\newcommand{\cm}[1]{}
\def\tNL{\tau\subs{NL}}
\def\fNL{f\subs{NL}}
\def\gNL{g\subs{NL}}

\def\fin{\phi}
\def\fcu{\sigma}
\def\Vi{V}
\def\Vc{U}
\def\vc{u}

\def\st{c}
\def\sh{*}
\def\se{e}

\def\sst{{}_{\st}{}}
\def\ssh{{}_{\sh}{}}
\def\sse{{}_{\se}{}}

\begin{document}
\title{Cosmological perturbations from a {\it Spectator} field during inflation}
\author{Lingfei Wang}
\author{and Anupam Mazumdar}
\affiliation{Consortium for Fundamental Physics, Lancaster University, Lancaster LA1 4YB, United Kingdom}

\abstract{In this paper we will discuss analytically the perturbations created from a slowly rolling subdominant spectator field which decays much before the end of inflation. The quantum fluctuations of such a spectator field can seed perturbations on very large scales and explain the temperature anisotropy in the cosmic microwave background radiation with moderate non-Gaussianity, provided the relevant modes leave the Hubble patch while the spectator is slowly rolling. Furthermore, the perturbations are purely {\it adiabatic} since the inflaton decay dominates and creates all the Standard Model degrees of freedom. We will provide two examples for the spectator field potential, one with a step function profile, and the other with an inflection point. In both the cases we will compute higher order curvature perturbations, i.e.\ local bispectrum and trispectrum, which can be constrained by the forthcoming Planck data.}
\maketitle

\secs{Introduction}
Inflation is the most well known paradigm for the early universe which has confronted the observations very well~\cite{Bennett:2012fp}. It is a dynamical mechanism which stretches the quantum perturbations on very large scales, while making the universe big and diluting  all matter~\cite{Guth:1980zm,Linde:1981mu,Albrecht:1982wi}. For a review on inflation models, see~\cite{Mazumdar:2010sa}. During inflation there could be more than one field participating in the dynamics -- they may drive inflation collectively as in the case of {\it assisted inflation}~\cite{Liddle:1998jc}, or there could be one or many subdominant fields who do not directly influence the overall dynamics.  Such a field might even decay early on while inflation is still going on, and we may regard it as a {\it spectator} field. In principle there could be many spectator fields \footnote{The spectator mechanism is different from the multi-field inflation scenarios discussed in \cite{Ashoorioon:2008qr,Battefeld:2011yj,Elliston:2011dr}.}.

If the potential for a spectator field is sufficiently flat before it decays, it can slow roll, and during this phase it can accumulate Hubble induced quantum fluctuations. These fluctuations can be imprinted in the temperature anisotropy of the cosmic microwave background (CMB) radiation. In the simplest scenario we may assume that the inflaton fluctuations are subdominant, and a single spectator field is present in the universe besides the inflaton which drives inflation.

The role of a subdominant field during inflation has been discussed before in the context of a curvaton scenario, where a light field seeds the fluctuations and then decays after inflation comes to an end~\cite{Enqvist:2001zp,Moroi:2001ct,Lyth:2001nq,Lyth:2002my}. Very recently we have discussed a novel possibility of creating perturbations from a spectator field which decays while inflation is still going on~\cite{Mazumdar:2012rs}. There are some crucial differences between these two scenarios which we highlight below.

Since inflation dilutes all matter, it is then necessary to create all the Standard Model degrees of freedom after the end of inflation, for a review on reheating, see \cite{Allahverdi:2010xz}. In the context of a curvaton scenario one possibility is that the curvaton produces all the relevant matter and dominates the energy density of the universe while decaying~\cite{Enqvist:2002rf,Enqvist:2003mr,Allahverdi:2006dr}. This is however difficult in most realistic scenarios, where we also require the curvaton to decay {\it solely} into the Standard Model degrees of freedom. Another possibility is that the curvaton subdominates the universe, so the relic isocurvature perturbations is weaker than the $5\%$ observational bound \cite{Bennett:2012fp}. Otherwise the decay products of the inflaton's and the curvaton's have to thermalize, but this is a daunting task. Usually implicit assumptions are being made that the decay products would thermalize, but there are few setups which can explicitly realize this in reality~\cite{Mazumdar:2011xe}. A subdominant curvaton also leads to a significant enhancement in non-Gaussianity~\cite{Lyth:2001nq,Lyth:2002my}. 

On the other hand if a field decays much before the end of inflation, it will never influence the thermal history of the universe and will become a {\it spectator} field. Its decay products will be redshifted away during inflation and the inflaton will be {\it solely} responsible for creating all the matter in the universe. Meanwhile such a spectator field could still be responsible for seeding the CMB anisotropy as discussed first in~\cite{Mazumdar:2012rs}, provided the relevant scales for the CMB leave the Hubble patch before the spectator field decays. In this respect there will be only {\it adiabatic} perturbations and {\it no isocurvature} perturbations in the universe.

In order for this scenario to work, obviously it is important that the spectator field's potential is very flat during inflation before it decays. There are no dearth of such fields, their origin could come from anywhere, i.e. from the visible sector (beyond the Standard Model sector whose fields carry the Standard Model charges), or from a hidden sector (fields which do not carry the Standard Model charges). Such a spectator does not even have to couple to the Standard Model degrees of freedom. All the onus will be now on the inflaton's coupling to the Standard Model degrees of freedom for creating the right thermal history of the universe. In this case inflation could be solely driven within the visible sector~\cite{Allahverdi:2006we,Allahverdi:2006iq}, or it may arise from any hidden sector with a specific coupling to the matter fields~\cite{Davidson:2000er,Allahverdi:2007zz,Feng:2013wn}. 

To be specific, we will derive general formulae for a generic potential of the inflaton and the spectator field, with the help of $\delta N$ formalism developed in \cite{Sasaki:1995aw,Wands:2000dp}, for a pedagogical discussion, see \cite{Lyth:2009zz}. Then we will consider an example where the spectator field has a sharp edge in the potential, where the slow roll conditions are naturally violated, and the final expressions are greatly simplified. In the second example we will consider a smooth transition from the slow roll phase to the oscillatory/decay phase of the spectator field. This can happen in the case where the spectator potential has an inflection point. The inflection point is where the second derivative of the potential vanishes but not the first or third derivatives. This gives a gentle slope for the spectator field to roll down the potential before the oscillation or the decay begins.

Before the spectator field decays, there exists an {\it entropy perturbations} due to the fluctuations in the spectator and in the inflaton field, although the inflaton's perturbations are subdominant.  Both the examples provide mild or large local bispectrum.  In the first example, the spectator has a flat and smooth potential, so the mild bispectrum is mostly sourced by the conversion from the short lived entropy perturbations to the curvature perturbations, which becomes non-Gaussian after the spectator decays. The second example has a curved potential for the spectator field, due to the inflection point, and therefore can generate significant local bispectrum from the potential curvature and from the conversion of the entropy perturbations into the curvature perturbations. In both cases, the non-Gaussianity is sourced by the entropy perturbations as it was highlighted in the context of non-perturbative decay of the inflaton~\cite{Jokinen:2005by,Enqvist:2004ey}. Eventually, the spectator field or its decay products are redshifted away during inflation, leaving no trace of isocurvature perturbations. In both the examples, the cosmological observables are predicted within our current observational limits. Note that this scenario generates non-Gaussianity in a different way than the multi-field inflation models, where large non-Gaussianities are expected due to non-trivial end-of-inflation boundary conditions, for example, see \cite{Mazumdar:2012jj}.

We also have to keep in mind that at the Hubble exit of the CMB's relevant scales, the spectator must have its classical slow roll dominating over its quantum fluctuations. Therefore the potential for the spectator field cannot be extremely flat or extremely subdominant. Otherwise this would give rise to a stochastic phase for the spectator field~\cite{Linde:2005ht}, which we will not explore here. This condition further puts a model independent lower bound on the energy density ratio between the spectator and the inflaton when the spectator decays. This lower bound will automatically prevent the generation of a very large local bispectrum or trispectrum.

In \refsec{General setup}, we will discuss the general setup. In \refsec{Curvature perturbations}, we will calculate the curvature perturbations at the linear and the higher orders. In \refsec{Plateau from a step function} and \refsec{Plateau from a saddle/inflection point}, we will provide examples for the spectator field with a step potential and an inflection point potential, respectively. We will conclude this paper in \refsec{Conclusion}. A list of variables are summarized in \refsec{Notations}.

\secs{General setup}
Let us consider a generic potential with the inflaton, $\phi$, and the spectator, $\sigma$. They don't have any interactions except their minimal couplings to gravity, so their potential has a general form:
\eq{ V\subs{tot}(\phi,\sigma)\equiv V(\phi)+ U(\sigma).\label{eq-in-Va}}
Here we assume both the fields are canonical scalar fields, which always satisfy $V(\phi)\gg U(\sigma)$. We also assume the spectator field $\sigma$ ends slow roll well before the end of inflation but after the Hubble exit of the pivot scale. This will give rise to two subsequent phases for $\sigma$ during inflation as shown in \refig{Timeline}:

\fig{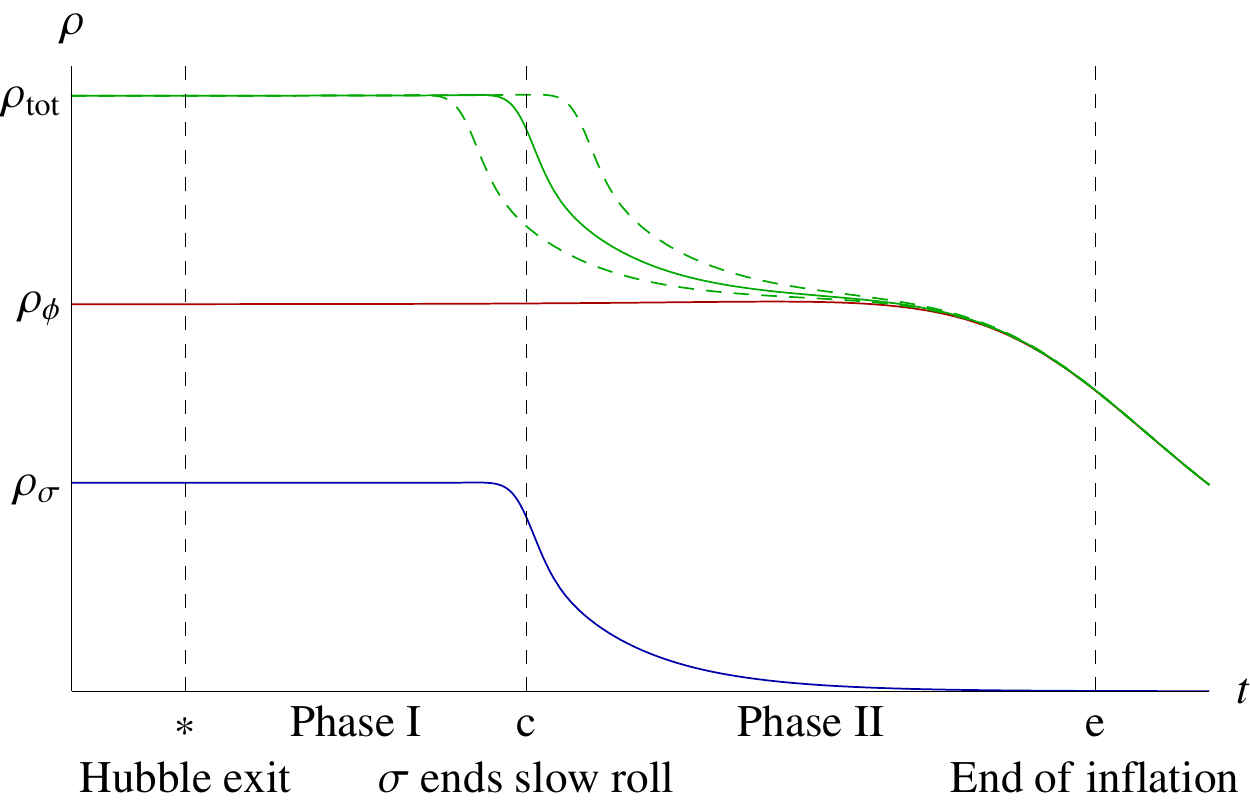}{A schematic timeline of the universe's evolution is shown above. The solid lines are the background evolutions of the energy densities of the inflaton, the spectator and the total contributions. The green dashed lines are the total energy densities of the universes with perturbed $\sigma$. The two phases of $\sigma$ evolution are separated by the phase boundary when $\sigma$ breaks the slow roll condition. The ``$*$'' denotes the epoch when $\sigma$ fluctuations leave the Hubble patch, ``$c$'' denotes the boundary between the two phases, and ``$e$'' denotes the end of inflation.}{0.8}{}

\begin{enumerate}
\item {\bf Phase I}: The inflaton $\phi$ leads inflation. Both $\phi$ and $\sigma$ are slowly rolling. This phase ends as the second order slow roll condition for $\sigma$ breaks down. We assume that the relevant perturbations for the CMB are leaving the Hubble patch in this phase~\footnote{ The observed pivot scale actually has a width of several e-folds. Here we consider every mode separately so the window is not shown in \refig{Timeline}.}.

\item {\bf Phase II}: When $\sigma$ ends slow roll (depicted here by point ``$c$''), the inflaton $\phi$ still dominates inflation under slow roll. Then $\sigma$ either oscillates around the minimum of it's potential, or decays instantly once its slow roll has been terminated. In either case, $\sigma$ or its decay products are being redshifted away swiftly during this phase, and can be regarded as a perfect fluid with a constant equation of state $w$. For this reason, several e-folds after the beginning of this phase, the dynamics reduces to that of a single field inflation.
\end{enumerate}

We assume that the primary curvature perturbations are generated from the $\sigma$ field and the inflaton's perturbations are negligible. During phase I, the energy density of $\sigma$ contributes to the Hubble rate, albeit small. But in phase II, its energy density or its decay product's contribution is quickly redshifted away by the ongoing inflationary expansion. Perturbations of the $\sigma$ field which are depicted in \refig{Timeline} by the dashed lines induce fluctuations of the phase boundary itself, which converts the $\sigma$ perturbations into curvature perturbations. 

The cosmological observables can then be calculated using the $\delta N$ formalism \cite{Sasaki:1995aw,Wands:2000dp}. According to $\delta N$ formalism we can regard separate universe patches as being perturbed as a whole by the perturbation modes at the Hubble exit. The different initial field perturbations $\delta\sigma{}_*{}$ at different universe patches then lead to different $\delta N$, the perturbations in the number of \emph{remaining} e-folds of inflation, which directly link to the amount of curvature perturbations as defined by $\zeta$~\cite{Bardeen:1983qw}, for a review see \cite{Mukhanov:1990me}.

For modes leaving the Hubble patch during phase I, the fluctuations, $\delta\fcu\ssh$, of a flat enough spectator field are Gaussian random perturbations with the power spectrum
\eq{P_{\delta\fcu\ssh}=\frac{H_\ast^2}{4\pi^2},}
where $H$ is the Hubble expansion rate of inflation, and ``$*$'' indicates the Hubble exit corresponding to the pivot scale. Imagining $N$ as a function of $\fcu\ssh$, we can write its perturbation as
\eq{\delta N=N_{\fcu}\delta\fcu\ssh+\frac{1}{2}N_{\fcu\fcu}(\delta\fcu\ssh)^2+\frac{1}{6}N_{\fcu\fcu\fcu}(\delta\fcu\ssh)^3+O((\delta\fcu\ssh)^4),\label{eq-in-dN}}
where $N_{\fcu}\equiv\partial N/\partial\fcu\ssh$, $N_{\fcu\fcu}\equiv\partial^2 N/\partial\fcu_*^2$ and $N_{\fcu\fcu\fcu}\equiv\partial^3 N/\partial\fcu_*^3$. Therefore the power spectrum for the curvature perturbations can be expressed at the first order as
\eq{P_\zeta=P_{\delta N}=N_{\fcu}^2P_{\delta\fcu\ssh}.}

The deviation from a Gaussian distribution of curvature perturbations is expressed order by order. The two lowest order parameters, namely the local bi-spectrum ($\fNL$) and the tri-spectrum ($\gNL$), have their relative local strengths characterized by~\footnote{ The other parameter for the strength of trispectrum, $\tNL$, is reduced to be proportional to $\fNL^2$, when the quantum fluctuations of one field dominate over the rest.}
\eq{\fNL=\frac{5}{6}\frac{N_{\fcu\fcu}}{N_{\fcu}^2},\hspace{0.6in}\gNL=\frac{25}{54}\frac{N_{\fcu\fcu\fcu}}{N_{\fcu}^3}.\label{eq-in-fgnl}}
Therefore in order to obtain the observational predictions, we need to work out the derivatives of $N$ as in \refeq{in-dN}. In accordance with the $\delta N$ formalism, we will use $\dd N=-H\dd t$.

\secs{Curvature perturbations}
In order to understand the effects of the perturbation $\delta\fcu\ssh$, we will consider the two phases separately and study how their numbers of e-folds are perturbed. After finishing the calculation at first order, we will use the results to obtain answers for the higher orders. Note that in our perturbative calculation the \emph{only initial} perturbation is $\delta\fcu\ssh$, while all the other $\delta$'s are the \emph{induced} perturbations by the initial perturbation $\delta\fcu\ssh$.

\ssecs{Perturbations from phase I}
We start from the background equations of motion for both the fields with the total potential \refeq{in-Va}. Under the slow roll conditions, we use $N$ as the remaining number of e-folds of inflation as a proper time, and rewrite their equations of motion as \footnote{ Our formalism for phase I is similar to Ref \cite{Vernizzi:2006ve}. However this formalism cannot be applied to phase II directly.}
\eq{\frac{\dd\fin}{\dd N}=\frac{\mpl^2\Vi'}{8\pi(\Vc+\Vi)},\hspace{0.5in}\frac{\dd\fcu}{\dd N}=\frac{\mpl^2\Vc'}{8\pi(\Vc+\Vi)},}
where primes on the potentials denote derivatives w.r.t the corresponding fields, i.e. $V'=\partial V/\partial\phi$ and $U'=\partial U/\partial\sigma$. These indicate a simple relation
\eq{\frac{\dd\fin}{\Vi'}=\frac{\dd\fcu}{\Vc'},}
whose integrated form is
\eq{\int_\fin^{\fin\ssh}\frac{\dd\fin}{\Vi'}=\int_\fcu^{\fcu\ssh}\frac{\dd\fcu}{\Vc'}.\label{eq-c11-slint}}

In the presence of an initial field perturbation $\delta\fcu\ssh$, \refeq{c11-slint} will also be perturbed, which leads to the perturbations of $\delta\fin$ and $\delta\fcu$ at any time, with the relation
\eq{-\frac{\delta\fin}{\Vi'}=\frac{\delta\fcu\ssh}{\Vc\ssh'}-\frac{\delta\fcu}{\Vc'}.\label{eq-c11-slpert}}

The phase boundary between phase I and phase II is the breakdown of the second order slow roll condition for $\fcu$
\eq{f\equiv\mpl^2\Vc''+8\pi(\Vc+\Vi),\label{eq-c11-f}}
so for the case we consider, at the phase boundary ``$\st$'' we will have $f\sst=0$. Since the phase boundary relation remains the same regardless of how the separate universe is perturbed, in the perturbed universe we would have the field values at the phase boundary also perturbed by the amount $\delta\fin\sst$ and $\delta\fcu\sst$, so it remains on the phase boundary
\eq{\delta f\bigl|_c=f_{\fin}\sst\delta\fin\sst+f_{\fcu}\sst\delta\fcu\sst=0,\label{eq-c11-boundp}}
where
\eq{f_{\fin}\sst\equiv\frac{\partial f}{\partial\fin}\biggl|_c=8\pi\Vi\sst',\hspace{0.5in}f_{\fcu}\sst\equiv\frac{\partial f}{\partial\fcu}\biggl|_c=\mpl^2\Vc\sst'''+8\pi\Vc\sst'.}
Taking \refeq{c11-slpert} at the phase boundary ``$c$'', together with \refeq{c11-boundp}, we are able to derive the induced field perturbations on the phase boundary by the initial perturbation $\delta\fcu\ssh$
\eq{\delta\fin\sst=-\frac{\Vi\sst'}{\Vc\ssh'}(1-\theta)\delta\fcu\ssh,\hspace{0.5in}\delta\fcu\sst=\frac{\Vc\sst'}{\Vc\ssh'}\theta\delta\fcu\ssh,}
where
\eq{\theta\equiv\frac{\Vi\sst'f_{\fin}\sst}{\Vi\sst'f_{\fin}\sst+\Vc\sst'f_{\fcu}\sst}=\frac{\epsilon_{\fin}\sst}{\epsilon_{\fin}\sst+\epsilon_{\fcu}\sst+\xi_{\fcu}\sst}\approx\frac{\epsilon_{\fin}\sst}{\xi_{\fcu}\sst}\ll1.}
Here
\eq{\xi_{\fcu}\equiv\frac{\mpl^4\Vc'\Vc'''}{64\pi^2(\Vc+\Vi)^2}\label{eq-c11-xic}}
is the third order slow roll parameter for $\fcu$ and is typically of order 1 or bigger for the case we consider. It can even be much greater than unity near the sharp edges of the spectator potential. On the other hand, the first order slow roll parameters for both the fields are
\eq{\epsilon_{\fin}\equiv\frac{\mpl^2\Vi'{}^2}{16\pi(\Vc+\Vi)^2}\ll1,\hspace{0.5in}\epsilon_{\fcu}\equiv\frac{\mpl^2\Vc'{}^2}{16\pi(\Vc+\Vi)^2}\ll1.}

Starting from ``$\sh$'', the Hubble exit of the mode we are interested in, the number of e-folds in phase I can be written as an integrated form on the uniform $\fin$ slicing
\eq{N_1\equiv\int_{N\sst}^{N\ssh}\dd N=\int_{\fin\sst}^{\fin\ssh}\frac{8\pi(\Vc+\Vi)}{\mpl^2\Vi'}\dd\fin.\label{eq-c11-N1-1}}
The perturbation in $N_1$ then becomes
\eq{\delta N_1=\frac{8\pi}{\mpl^2}\left(-\frac{\Vc\sst+\Vi\sst}{\Vi\sst'}\delta\fin\sst+\int_{\fin\sst}^{\fin\ssh}\frac{\Vc'}{\Vi'}\delta\fcu\dd\fin\right).\label{eq-c11-N1-2}}
On uniform $\fin$ slicing we always have $\delta\fin=0$, so \refeq{c11-slpert} becomes $\delta\fcu=\Vc'\delta\fcu\ssh/\Vc\ssh'$. This simplifies the integral in \refeq{c11-N1-2} to
\eq{\int_{\fin\sst}^{\fin\ssh}\frac{\Vc'}{\Vi'}\delta\fcu\dd\fin=\frac{\delta\fcu\ssh}{\Vc\ssh'}\int_{\fin\sst}^{\fin\ssh}\frac{\Vc'{}^2}{\Vi'}\dd\fin=\frac{\Vc\ssh-\Vc\sst}{\Vc\ssh'}\delta\fcu\ssh.}
Consequently, we get the final expression for the perturbed number of e-folds in phase I as
\eq{\delta N_1=\frac{8\pi\bigl(\Vc\ssh-\Vc\sst+(1-\theta)(\Vc\sst+\Vi\sst)\bigr)}{\mpl^2\Vc\ssh'}\delta\fcu\ssh.\label{eq-c11-dN1}}

\ssecs{Perturbations from phase II}
As explained in \refsec{General setup}, during phase II we have a single field inflation with an additional perfect fluid component whose energy density came from either the decay products or the oscillations of $\sigma$. The Hubble expansion rate is given by
\eq{H^2=\frac{8\pi}{3\mpl^2}\left(\Vc\sst e^{3(1+w)(N-N\sst)}+\Vi(\fin)\right)=\frac{\Vi'(\fin)}{3}\frac{\dd N}{\dd\fin},\label{eq-c12-Hubble}}
where the second half comes from the slow roll equation of motion for $\phi$. Here $U_c$ means the value of $U(\sigma)$ at the phase boundary $c$, and $w$ is the equation of state for the perfect fluid.

It is obvious that \refeq{c12-Hubble} is actually a first order differential equation between $N$ and $\fin$. Its exact solution is given by
\eq{N_2\equiv N\sst-N\sse=n(\fin\sst,\fin\sse)+\frac{1}{3(1+w)}\ln\frac{1-\alpha r}{1-r},\label{eq-c12-N2}}
where
\eq{n(\fin_1,\fin_2)\equiv\int_{\fin_2}^{\fin_1}\frac{8\pi\Vi}{\mpl^2\Vi'}\dd\fin\label{eq-c12-n}}
is the number of e-folds of inflation when $\fin$ serves as the \emph{only} component of the universe, while the second term is the contribution from the perfect fluid. In \refeq{c12-N2} we have defined
\eq{r\equiv\frac{\Vc\sst}{\Vc\sst+\Vi\sst}\ll1\label{eq-c12-r}}
as the energy density ratio of the perfect fluid at the phase boundary, and
\eq{\alpha\equiv1-\frac{24(1+w)\pi\Vi\sst}{\mpl^2}\int_{\fin\sse}^{\fin\sst}\frac{e^{-3(1+w)n(\fin\sst,\fin)}}{\Vi'(\fin)}\dd\fin\ll1\label{eq-c12-alpha}}
will be calculated in \refsec{The alpha}. Note that $\alpha$ is order of the slow roll parameter $\epsilon_{\phi c}\ll1$, and hence $\alpha\ll1$.

In a perturbed universe with $\delta\fcu\ssh$, field perturbations $\delta\fin\sst$ and $\delta\fcu\sst$ will be generated on the phase boundary. They will give rise to the perturbations of the above parameters
\eq{\delta n(\fin\sst,\fin)=\frac{8\pi\Vi\sst}{\mpl^2\Vi\sst'}\delta\fin\sst,}
\eq{\delta r=\frac{1}{\Vc\sst+\Vi\sst}\bigl((1-r)\Vc\sst'\delta\fcu\sst-r\Vi\sst'\delta\fin\sst\bigr),}
and
\eq{\delta\alpha=-\left((1-\alpha)\frac{\Vi\sst'}{\Vi\sst}+\alpha\frac{24(1+w)\pi\Vi\sst}{\mpl^2\Vi\sst'}\right)\delta\fin\sst.}
Feeding them back into \refeq{c12-N2} gives the perturbation in the number of e-folds for phase II
\eq{\delta N_2=\frac{1}{1-\alpha r}\left(\frac{8\pi\Vi\sst}{\mpl^2\Vi\sst'}\delta\fin\sst+\frac{(1-\alpha)\Vc\sst'}{3(1+w)(\Vc\sst+\Vi\sst)}\delta\fcu\sst\right).\label{eq-c12-dN2}}

\ssecs{Total perturbations}
Adding up \refeq{c11-dN1} and \refeq{c12-dN2} gives the total perturbation in the number of e-folds generated by the initial perturbation $\delta\fcu\ssh$
\eq{\delta N=\delta N_1+\delta N_2=N_{\fcu}\delta\fcu\ssh,}
where
\eq{N_{\fcu}=\frac{1-\alpha}{1-\alpha r}\left(\frac{8\pi\Vc\ssh}{\mpl^2\Vc\ssh'}+\theta\Bigl(\frac{\Vc\sst'{}^2}{3(1+w)(\Vc\sst+\Vi\sst)\Vc\ssh'}-\frac{8\pi\Vc\sst}{\mpl^2\Vc\ssh'}\Bigr)\right)+\frac{\alpha(1-r)}{1-\alpha r}\frac{8\pi(\Vc\ssh-\Vc\sst)}{\mpl^2\Vc\ssh'}.\label{eq-c1t-Np1}}
When the potential $\Vc(\sigma)$ is sharp enough at the phase boundary, the conditions $\xi_{\fcu}\sst\gg\epsilon_{\fin}\sst$ and $r\xi_{\fcu}\sst\gg\epsilon_{\fin}\sst\epsilon_{\fcu}\sst$ are satisfied. This is the case we will be analyzing in this paper. Under these conditions, \refeq{c1t-Np1} is dominated by the very first term. It will simplify to
\eq{N_{\fcu}\approx\frac{8\pi\Vc\ssh}{\mpl^2\Vc\ssh'}.\label{eq-c1t-Np}}
We see that \refeq{c1t-Np} comes in agreement with the results obtained in Ref.~\cite{Mazumdar:2012rs}. When the spectator dominates the curvature perturbations, we need the spectator's potential should be relatively flatter than that of the inflaton's \footnote{\ This is because when the inflaton totally dominates the curvature perturbations, we have a similar expression $N_\fin=8\pi \Vi\ssh/\mpl^2\Vi\ssh'$.}
, i.e.\ $\Vc\ssh/\Vc\ssh'\gg\Vi\ssh/\Vi\ssh'$.

The above calculations would then give the power spectrum of the curvature perturbations
\eq{P_\zeta=\frac{16H_*^2\Vc_*^2}{\mpl^4\Vc\ssh'{}^2}=\frac{H_*^2s^2r^2}{\pi s_{\fcu}^2\mpl^2\epsilon_{\fcu}\ssh},\label{eq-c1t-Pz}}
where
\eq{s\equiv\frac{\Vc\sst+\Vi\sst}{\Vc\ssh+\Vi\ssh}<1}
is the ratio between the total energy densities at the phase boundary and at the Hubble exit.

The spectral tilt is given by~\footnote{ The $-2\epsilon_{\fcu}\ssh$ term is neglected because it is much smaller than the last term in \refeq{c1t-ns0}.}
\eq{n_s-1\equiv-\frac{\partial\ln P_\zeta}{\partial N}=-2\epsilon_{\fin}\ssh+2\eta_{\fcu}\ssh-\frac{4s_{\fcu}\epsilon_{\fcu}\ssh}{sr},\label{eq-c1t-ns0}}
where
\eq{s_{\fcu}\equiv\frac{\Vc\sst}{\Vc\ssh}<1}
is the energy density ratio of $\fcu$ between the phase boundary and the Hubble exit.

Note from \refeq{c1t-Pz}, we can obtain a constraint on $\epsilon_{\fcu}\ssh$ from the observed $P_\zeta$, leading to
\eq{\epsilon_{\fcu}\ssh=\frac{H_*^2s^2r^2}{\pi s_{\fcu}^2\mpl^2P_\zeta}.\label{eq-c1t-epssx}}
By plugging \refeq{c1t-epssx} back into \refeq{c1t-ns0}, we see that for the inflation energy scale $H\ssh<10^{-5}\mpl$, the last term in \refeq{c1t-ns0} is always negligible compared to the observed spectral index, $n_s\approx0.96$~\cite{Bennett:2012fp}. So we are left with
\eq{n_s-1=-2\epsilon_{\fin}\ssh+2\eta_{\fcu}\ssh.\label{eq-c1t-ns}}

With the help of \refeq{c1t-ns0}, the running of spectral tilt can be shown as
\eq{\frac{\dd n_s}{\dd\ln k}=-\frac{\dd n_s}{\dd N}=-\frac{1}{2}(n_s-1)^2-2\Bigl(3\epsilon_*^2-2(\epsilon_{\fin}\ssh\eta_{\fin}\ssh+\epsilon_{\fcu}\ssh\eta_{\fcu}\ssh)-\eta_{\fcu*}^2+\xi_{\fcu}\ssh\Bigr),\label{eq-c1t-nsr}}
where
\eq{\epsilon\equiv\frac{\dd}{\dd t}\frac{1}{H}=\epsilon_{\fin}+\epsilon_{\fcu}}
is the total slow roll parameter for inflation, and $\epsilon\ssh\approx\epsilon_{\fin}\ssh$ can be taken in \refeq{c1t-nsr} for  $H\ssh<10^{-5}\mpl$.

\ssecs{Higher order perturbations}
Higher order curvature perturbations, i.e.\ the local bispectrum, $\fNL$, and the trispectrum, $\gNL$, require the expression of $\alpha$, which is derived in \refsec{The alpha}.

The strength of the local bispectrum $\fNL$ can be derived from taking the derivative $\partial/\partial\fcu\ssh$ on \refeq{c1t-Np1}. The leading terms are~\footnote{ We take series expansion for the parameters $\epsilon_{\fin}\ssh\ll1$, $\epsilon_{\fin}\sst\ll1$, $|\eta_{\fin}\ssh|\ll1$, $|\eta_{\fin}\sst|\ll1$, $\epsilon_{\fcu}\ssh\ll1$, $\epsilon_{\fcu}\sst\ll1$, $|\eta_{\fcu}\ssh|\ll1$, $r\ll1$ and $\xi_{\fin c}$, by writing down only the leading order contributions for the expressions of $\fNL$ and $\gNL$. The slow roll parameter $\xi_{\fin}$ does not have to be much smaller than 1, but in most cases it has the order of $\epsilon_{\fin c}^2$. Therefore here we also take it as a small quantity.}
\eqa{\fNL&\equiv&\frac{5}{6}\frac{N_{\fcu\fcu}}{N_{\fcu}^2}=\frac{10s_{\fcu}^2\epsilon_{\fin}\sst}{3r}\left(\frac{2\epsilon_{\fin}\sst-\eta_{\fin}\sst}{3(1+w)}-\frac{\epsilon_{\fin}\sst}{\xi_{\fcu c}^2}\biggl(1-\frac{\lambda_{\fcu}\sst}{\xi_{\fcu}\sst}\biggr)+\frac{\eta_{\fin}\sst}{\xi_{\fcu}\sst}\right)\nonumber\\
&+&\frac{20s_{\fcu}^2\epsilon_{\fcu}\sst\epsilon_{\fin}\sst}{9(1+w)r^2\xi_{\fcu c}}\left(2-\frac{4+3w}{\xi_{\fcu}\sst}-\frac{\lambda_{\fcu}\sst}{\xi_{\fcu c}^2}\right)+\frac{5s_{\fcu}^2\epsilon_{\fcu}\ssh}{3s^2r^2}-\frac{5s_{\fcu}\eta_{\fcu}\ssh}{6sr}\nonumber\\
&+&\mathrm{higher\ order\ terms},\label{eq-c2-Nppr1}}
where the third and fourth order slow roll parameters $\xi_{\fcu}$ and $\lambda_{\fcu}$ are defined as~\footnote{ Note that the third and higher orders slow roll parameters of $\fcu$ (e.g.\ $\xi_{\fcu}$ and $\lambda_{\fcu}$) typically vary from $O(1)$ to (positive or negative) infinity depending on the actual potential $U(\sigma)$. Therefore we do not take them as infinitesimals and do not take series expansion on them in \refeq{c2-Nppr1} as well as in \refeq{ap2-Nppp}.}
\eq{\xi_{\fcu}\equiv\frac{\mpl^4\Vc'\Vc'''}{(8\pi(\Vc+\Vi))^2},\hspace{0.4in}\lambda_{\fcu}\equiv\frac{\mpl^6\Vc'{}^2\Vc''''}{(8\pi(\Vc+\Vi))^3}.\label{eq-c2-lc}}

The third order derivative $N\subs{\fcu\fcu\fcu}=\partial^2N_{\fcu}/\partial\fcu_*^2$ can be calculated in the same way and also $\gNL$. According to \refeq{in-fgnl}, we obtain the leading order trispectrum of curvature perturbations as
\eq{\gNL=\frac{25s_{\fcu}^2}{54r^2}\left(\frac{2\eta_{\fcu}\ssh}{s^2}\biggl(\eta_{\fcu}\ssh-\frac{s_{\fcu}\epsilon_{\fcu}\ssh}{sr}\biggr)-\frac{\xi_{\fcu}\ssh}{s^2}+4s_{\fcu}\epsilon_{\fin}\sst\xi_{\fin}\sst\biggl(\frac{1}{3(1+w)}-\frac{1}{\xi_{\fcu}\sst}\biggr)+\frac{8s_{\fcu}\epsilon_{\fin}\sst\epsilon_{\fcu}\sst}{3(1+w)r\xi_{\fcu}\sst}A\right),\label{eq-ap2-Nppp}}
where
\eqa{A&\equiv&\xi_{\fin}\sst+\eta_{\fin}\sst\biggl(2\eta_{\fin}\sst+\frac{3\eta_{\fcu}\ssh}{s}\biggr)-3\epsilon_{\fin}\sst\biggl(2\eta_{\fin}\sst+\frac{\eta_{\fcu}\ssh}{s}\biggr)\biggl(2-\frac{4+3w}{\xi_{\fcu}\sst}-\frac{\lambda_{\fcu}\sst}{\xi_{\fcu c}^2}\biggr)\nonumber\\
&+&2\epsilon_{\fin c}^2\biggl(6-\frac{2}{\xi_{\fcu}\sst}-\frac{3(1+w+\lambda_{\fcu}\sst)}{\xi_{\fcu c}^2}+\frac{3(4+3w)\lambda_{\fcu}\sst-\chi_{\fcu}\sst}{\xi_{\fcu c}^3}+\frac{3\lambda_{\fcu c}^2}{\xi_{\fcu c}^4}\biggr),}
and
\eq{\chi_\sigma\equiv\frac{\mpl^8\Vc'{}^3\Vc'''''}{(8\pi(\Vc+\Vi))^4}}
is the fifth order slow roll parameter for $\sigma$.

Here we pause and discuss the validity of the perturbation theory itself. In all the above calculations, we have assumed that the classical slow roll dominates over the quantum fluctuations of $\sigma$ for the relevant scales. Here we focus on comparing the strengths of quantum fluctuations and classical slow roll, ensuring the validity of the above calculations.

A dominant classical slow roll requires the field displacement of $\fcu$ to be larger than the quantum fluctuations in one Hubble time, i.e. $H_*/2\pi<d\sigma/H_*dt$, or
\eq{P_{\delta\fcu\ssh}<\left(\frac{\dd\fcu}{\dd N}\right)^2.} 
Multiplying both sides by $N_{\fcu}^2$, we convert the l.h.s into the power spectrum of the curvature perturbations $P_\zeta$. A simple exercise then yields a \emph{model independent} lower bound on $r$
\eq{r>\frac{s_{\fcu}}{s}\sqrt{P_\zeta}.\label{eq-vp-rl2}}
Since from observations $P_\zeta\approx2.5\times10^{-9}$, we obtain a lower bound $r>5\times10^{-4}$ from a simple estimation by using $s_{\fcu}\equiv\Vc\sst/\Vc\ssh\approx1$ and $s\equiv(\Vc\sst+\Vi\sst)/(\Vc\ssh+\Vi\ssh)\approx0.1$.

From \refeq{c2-Nppr1} and \refeq{ap2-Nppp}, we see that $\fNL$ and $\gNL$ depends very much on the magnitude of $r$. Therefore a lower bound on $r$ typically becomes an upper bound for $\fNL$ and $\gNL$. In many cases this constraint automatically prevents the possibility of a very large non-Gaussianity.

\section{A step potential for the spectator field\label{sec-Plateau from a step function}}
\ssecs{General predictions}

Since the spectator field decays during inflation and all the matter are created by the inflaton, we illustrate a very simple potential for $U(\sigma)$, which is given by
\eq{\Vc(\fcu)=\mathrm{Step}(\fcu-\fcu_0)\vc(\fcu)=\left\{\begin{array}{l@{\hspace{0.3in}}l}\vc(\fcu),&\mathrm{for\ }\fcu>\fcu_0,\\0,&\mathrm{else},\end{array}\right.\label{eq-il-U0}}
where we assume $\vc(\fcu)$ is flat and smooth enough to accommodate slow roll for $\sigma>\sigma_0$. The step function is a limiting case of the hyperbolic tangent function
\eq{\mathrm{Step}(x)=\lim_{k\rightarrow+\infty}\frac{1+\tanh kx}{2}.}

We also assume that in the early universe $\fcu$ would be initially displaced at $\fcu>\fcu_0$, slowly rolling down the potential $\vc(\fcu)$. The field $\fcu$ ends slow roll at $\fcu\sst\approx\fcu_0$, because of the violation of the second order slow roll condition. At this point its energy density is transferred to the ideal fluid with a constant equation of state $w$. On the other hand when $\fcu$ ends slow roll, it still remains on the $u(\sigma)$ potential which ensures $\tanh k(\fcu\sst-\fcu_0)\approx1$, i.e.\ $k(\fcu\sst-\fcu_0)\gg1$. Therefore the derivatives of the $\tanh$ function at the point $\fcu\sst$ has a leading contribution
\eq{\frac{\partial^n\tanh k(\fcu-\fcu_0)}{\partial\fcu^n}\approx(-2)^{n+1}k^ne^{-2k(\fcu-\fcu_0)},\hspace{0.3in}\mathrm{for\ }n=1,2,\dots\mathrm{\ under\ }k(\fcu-\fcu_0)\gg1.}

With the phase boundary condition $f=0$ from \refeq{c11-f}, we find the deviation of $\fcu\sst$ from $\fcu_0$ as
\eq{\fcu\sst-\fcu_0=\frac{1}{2k}\ln\frac{k^2\mpl^2\Vc\sst}{2\pi(\Vc\sst+\Vi\sst)}.}
The higher order slow roll parameters then become
\eqa{\xi_{\fcu}\sst&=&2k\mpl\sqrt\frac{\epsilon_{\fcu}\sst}{\pi},\\
\lambda_{\fcu}\sst&=&-2k^2\mpl^2\frac{\epsilon_{\fcu}\sst}{\pi},\\
\chi_{\fcu}\sst&=&2k^3\mpl^3\left(\frac{\epsilon_{\fcu}\sst}{\pi}\right)^\frac{3}{2}.}

For a step function potential whose edge is infinitely sharp (i.e. $k\rightarrow\infty$), and the potential $u(\sigma)$ is very flat and smooth (i.e.\ $\epsilon_{\sigma*}\ll1$, $\eta_{\sigma*}\ll1$, and $s_\sigma\approx1$), the local bispectrum \refeq{c2-Nppr1} is dominated by the first term, where we also have $\xi_{\sigma c}\rightarrow\infty$. The bispectrum and the trispectrum are therefore given by
\eqa{\fNL&\approx&\frac{10\epsilon_{\fin}\sst(2\epsilon_{\fin}\sst-\eta_{\fin}\sst)}{9(1+w)r},\label{eq-il-Npp2}\\
\gNL&\approx&-\frac{25\xi_{\fcu}\ssh}{54s^2r^2}+\frac{50\epsilon_{\fin}\sst\xi_{\fin}\sst}{81(1+w)r^2}.\label{eq-il-Nppp2}}
From \refeq{il-Npp2} we find an agreement with \cite{Mazumdar:2012rs}.

\ssecs{Double quadratic potentials}
As a naive example, we consider both fields to have quadratic potentials. The potential $\Vc(\fcu)$ also has a step function for a sharp transition during inflation. In this case, the potentials are written as~\footnote{ Because $\fcu$ rolls very slowly during phase I, we typically expect $\fcu\ssh-\fcu_0\ll\fcu_0$. This means in phase I the effective potential for $\fcu $ is actually $\Vc(\fcu)=m_{\fcu}^2\fcu_{0}^2+2m_{\fcu}^2\fcu_0(\fcu-\fcu_0)$.}
\eq{\Vi(\fin)=m_{\fin}^2\fin^2,\hspace{0.5in}\Vc(\fcu)=m_{\fcu}^2\fcu^2\,\mathrm{Step}(\fcu-\fcu_0).}

Inflation is driven by $V(\phi)$. In the beginning when the relevant perturbations leave the Hubble patch, we have $\fcu\ssh>\fcu_0$, so it stays on the plateau and rolls down slowly. When the spectator field reaches $\fcu\sst=\fcu_0$, the sudden change in the potential would break the second order slow roll condition for $\fcu$, ends its slow roll, and let it decay instantly to radiation as we assume. The radiation is quickly diluted away by inflation.

\fig{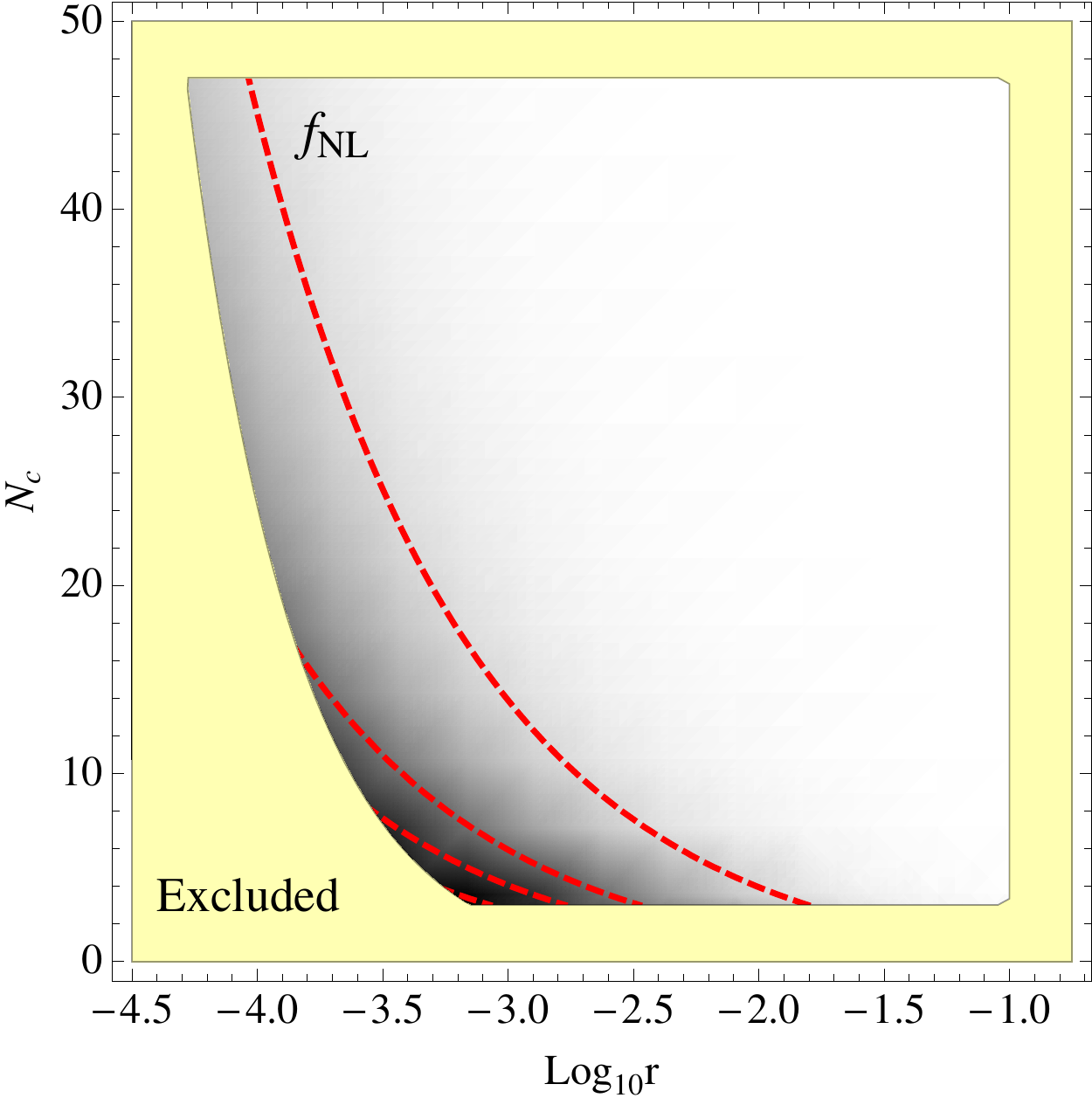}{The local bispectrum strength $\fNL$ is shown for a spectator field with a step potential. The yellow shaded region is excluded in the parameter space as discussed by five points in \refsec{Plateau from a step function}. The darker regions indicate a higher $\fNL$. The red dashed contours are for $\fNL=1,5,10,20$ from the top right to the bottom left.}{0.8}{h}

For this model, we have a total of 4 free parameters --- $m_{\fin},m_{\fcu},\fcu_0$, and $N\sst$ -- the number of e-folds from the phase boundary ``$\st$'' to the end of inflation. The overall energy scale only affects the power spectrum of curvature perturbation by
\eq{P_\zeta=\frac{16(2N\ssh+1)m_\phi^2\sigma_0^2}{3\mpl^4}.}
Therefore we can fix $P_\zeta$ to the observed value and hence reduce the number of parameters to 3. We want $\fcu$ to dominate the curvature perturbations, which requires $\fcu_0$ to be large enough ($\fcu_0\gg\fin\ssh$). However as long as this condition is satisfied, the value of $\fcu_0$ hardly changes the cosmological predictions of the model. We are left with only two free parameters which are the mass ratio $m_{\fcu}/m_{\fin}$ and $N\sst$. For this model the energy density ratio at the phase boundary is given by
\eq{r=\frac{m_{\fcu}^2\fcu_0^2}{m_{\fin}^2\fin\sst^2},}
so we will use $r$ instead of $m_{\fcu}/m_{\fin}$ for the parameter space, together with $N\sst$.

After transforming the parameter space, the free parameters reduce from $m_{\fin},m_{\fcu},\fcu_0,N\sst$ to $r$ and $N\sst$. The other two degrees of freedom are absorbed in $P_\zeta$ which is fixed by observation, and $\fcu_0$ which does not change the cosmological predictions.\ \footnote{ In fact $\fcu_0$ changes the inflation energy scale and the ratio of the curvature perturbations by the inflaton and the spectator, but since for $\fcu_0\gg\phi\sst$, the spectator totally dominates the curvature perturbations and the inflation energy scale is too low to be seen from the primordial gravitational waves, its value does not matter any more.} The transformation in the parameter space has the following relations
\eqa{m_\phi^2&=&\frac{3\mpl^4P_\zeta}{16(2N\ssh+1)\sigma_0^2},\label{eq-il-mphi}\\
m_\sigma^2&=&\frac{3(2N_c+1)\mpl^6P_\zeta}{64(2N_*+1)\pi\fcu_0^4}.}
The local bispectrum in \refeq{il-Npp2} now simplifies to (after neglecting the $O(1)$ coefficient)
\eq{\fNL\sim\frac{1}{(2N_c+1)^2r}.}

The two dimensional parameter space for $r$ and $N\sst$ is constrained by the following conditions.
\begin{enumerate}
\item The Hubble exit of the pivot scale, the phase boundary and the end of inflation are all well separated. So typically we choose $3\le N\sst\le N\ssh-3$.
\item The $\fcu$ field should be subdominant, so $r\ll1$.
\item The inflaton should provide suppressed curvature perturbations compared to that of the spectator, which means $\sigma_0\gg\phi_*$.
\item The first order slow roll parameter is smaller than unity when $\fcu$ stays on the flat potential, i.e. $\epsilon_{\fcu}\ssh<1$ and $\epsilon_{\fcu}\sst<1$.
\item The quantum fluctuations of $\fcu$ should not dominate over its classical slow roll. This means \refeq{vp-rl2} is valid.
\end{enumerate}
Because of the last constraint, the energy density ratio $r$ can not be too small at the phase boundary. Therefore the maximum possible value $\fNL\sups{(max)}\sim20$ is achieved when $N_c=3$ and $r\sim10^{-3}$.

Under the above conditions, we can calculate the spectral index $n_s$, its running $\dd n_s/\dd\ln k$, the local bispectrum $\fNL$, and the trispectrum $\gNL$, according to \refeq{c1t-ns0}, \refeq{c1t-nsr}, \refeq{il-Npp2} and \refeq{il-Nppp2}. A specific example of $\fNL$ is shown in \refig{Step}, for the parameters $N\ssh=50$, $\fcu_0=10\mpl$ and $w=1/3$. The regions violating any of the above five conditions are excluded and shown by the yellow shaded region in \refig{Step}. We can read out from \refeq{il-mphi} the mass for $\phi$ here is $m_\phi\approx2.1\times10^{-7}\mpl$, so the curvature perturbations from $\fin$ are indeed negligible. In addition, for these parameters we have $n_s=0.98$, $\dd n_s/\dd\ln k=-3\times10^{-4}$ and $\gNL\ll1$, all of which hardly depend on the choice of $N\sst$ or $r$, and fall within the observational bound.

From \refig{Step}, we see that the parameter space is limited. In particular, $r$ is constrained on both the sides because we need $\fcu$ to be subdominant and its quantum fluctuations not to overcome the slow roll motion. Moreover, $\fcu$ hardly contributes to $n_s$, $\fNL$ or $\gNL$ because all of its slow roll parameters are tiny. With the inflaton $\fin$ being the only contribution, we get $n_s\approx0.98$, and small running, bispectrum and trispectrum. For these parameters, we can see the local bispectrum strength $\fNL$ indeed has a maximum value around 20. The parameter space for $\fNL>10$ is very limited. In this case the major contribution to $\fNL$ comes from the conversion of the entropy to the curvature perturbations, which becomes non-Gaussian after the spectator ends slow roll and decays into a perfect fluid, even though this non-Gaussian conversion only lasts for one e-fold or so before the entropy perturbations are redshifted away.

\section{An inflection point potential for the spectator field\label{sec-Plateau from a saddle/inflection point}}
\ssecs{Generic inflaton potential}
Flat directions naturally arise in string theory or supersymmetric theories \cite{Mazumdar:2010sa}. These flat directions can also be candidates for $\Vc(\fcu)$. In most cases, such flat directions can be written locally as an effective scalar potential in the form~\cite{Allahverdi:2006we,Allahverdi:2006iq}
\eq{\Vc(\Delta\fcu)=\Vc_0\left(1+\gamma_1\frac{\Delta\fcu}{\mpl}+\frac{\gamma_3}{6}\frac{\Delta\fcu^3}{\mpl^3}+O\biggl(\frac{\Delta\fcu^4}{\mpl^4}\biggr)\right),\label{eq-e-U}}
where
\eq{\Delta\fcu\equiv\fcu-\fcu_0.}
Therefore at $\Delta\fcu=0$, i.e.\ $\fcu=\fcu_0$, we will get an inflection/saddle point where $\Vc=\Vc_0$ and $\Vc''=0$, and $\Delta\fcu$ is the deviation from the inflection point $\sigma_0$. For inflection and saddle points we will have $\gamma_1>0$ and $\gamma_1=0$ respectively, and we always have $\gamma_3>0$. In general the higher order terms in the effective potential \refeq{e-U}, e.g.\ $\Delta\fcu^4$, also provide a small contribution to the potential or its derivatives. Here we assume their contribution vanishes for the sake of simplicity.

In this respect the motion of $\fcu$ can be solved as follows. We first obtain the deviation at the phase boundary $\Delta\fcu\sst$, from the breakdown of second order slow roll condition $\eta_{\fcu}\sst=-1$
\eq{\Delta\fcu\sst=-\frac{8\pi\mpl(U_c+V_c)}{\gamma_3 U_0}.\label{eq-sip-eta}}
With this we can introduce a very helpful parameter $\gamma_0$, which tells us how ``flat'' the potential is at the inflection point
\eq{\gamma_0\equiv\sqrt{\frac{\gamma_1}{\mpl}\frac{2\mpl^3}{\gamma_3\Delta\fcu\sst^2}}=\sqrt\frac{\gamma_1\gamma_3}{2}\,\frac{\Vc_0}{4\pi(\Vc\sst+\Vi\sst)}.}
Therefore the ratio of $U'(\sigma)$ between the inflection point and the phase boundary is $\gamma_0/(1+\gamma_0)$. For the inflection point potential we discuss here, typically $\gamma_0\ll1$.

As long as we specify the inflaton potential and $N\sst$, we are able to solve the equation of motion for $\Delta\fcu$. The slow roll approximation in phase I gives the l.h.s of \refeq{c11-slint} as
\eqa{\int_{\Delta\fcu\sst}^{\Delta\fcu\ssh}\frac{\dd\Delta\fcu}{\Vc'}&=&\frac{\mpl^2}{\Vc_0}\sqrt\frac{2}{\gamma_1\gamma_3}\left.\arctan\sqrt\frac{\gamma_3}{2\gamma_1}\,\frac{\Delta\fcu}{\mpl}\right|_{\st}^{\sh}\nonumber\\
&=&\frac{\mpl^2}{4\pi\gamma_0(\Vc\sst+\Vi\sst)}\left(\arctan\frac{1}{\gamma_0}+\arctan\frac{x\ssh}{\gamma_0}\right),\label{eq-sip-dc1}}
where $x\equiv\Delta\sigma/|\Delta\sigma_c|$ is the relative displacement from the inflection point.

Because $\fcu$ always subdominates the energy density, we can neglect its contribution to the Hubble rate and solve the background evolution for $\phi(N)$. By equating \refeq{sip-dc1} (as the l.h.s of \refeq{c11-slint}) with the r.h.s of \refeq{c11-slint}, we derive the evolution of $\sigma$ by $x_*(N_*)$
\eq{\arctan\frac{x_*}{\gamma_0}=-\arctan\frac{1}{\gamma_0}+\frac{4\pi\gamma_0(\Vc\sst+\Vi\sst)}{\mpl^2}\int_{\fin\sst}^{\fin(N_*)}\frac{\dd\fin}{\Vi'}.\label{eq-sip-dc2}}

We can also derive the slow roll parameters that are needed to calculate the cosmological observables
\eq{\epsilon_{\sigma}=\frac{64\pi^3(U_c+V_c)^4}{\gamma_3^2U_0^2(U+V)^2}(\gamma_0^2+x^2)^2,\hspace{0.3in}\eta_{\sigma}=\frac{U_c+V_c}{U+V}x,\hspace{0.3in}\xi_{\sigma}=\frac{(U_c+V_c)^2}{2(U+V)^2}(\gamma_0^2+x^2),}
while all the higher order slow roll parameters vanish. When $\gamma_3$ is large and $\gamma_0\ll1$, we have $s_\sigma\approx1$, and the final results are simplified to
\eqa{n_s-1&=&-2\epsilon_{\phi*}+2 x_*,\label{eq-sip-ns1}\\
\frac{\dd n_s}{\dd\ln k}&\approx&-\frac{1}{2}(n_s-1)^2+2\epsilon_{\phi*}(2\eta_{\phi*}-3\epsilon_{\phi*})+\gamma_0^2-x_*^2,\label{eq-sip-dns1}\\
\fNL&\approx&-\frac{10(5+6w)\epsilon_{\phi c}(2\epsilon_{\phi c}-\eta_{\phi c})}{9(1+w)r}-\frac{5x_*}{6r},\label{eq-sip-fnl1}\\
\gNL&\approx&-\frac{50(5+6w)\epsilon_{\phi c}\xi_{\phi c}}{81(1+w)r^2}+\frac{25(3x_*^2-\gamma_0^2)}{108r^2}.\label{eq-sip-gnl1}}

Typically, we have $|x_*|\sim O(10^{-2})$, and $\epsilon_{\phi c}\sim\eta_{\phi c}\lesssim 10^{-1}$. For this reason in \refeq{sip-fnl1}, the $x_*$ term dominates in general, and becomes the major contribution to $\fNL$. The $x_*$ term comes from $\eta_{\fcu*}$ in the original equation \refeq{c2-Nppr1}. A brief estimation for the maximum local bispectrum can be achieved for $|x_*|\sim10^{-2}$ with a small $r\sim10^{-4}$, so $\bigl|\fNL\sups{(max)}\bigr|\sim100$.

\ssecs{Quadratic slow roll inflaton}
As a simple example, we consider the inflaton potential to be
\eq{\Vi(\fin)=m^2\fin^2,\label{eq-sipq-V}}
and we assume it dominates over the spectator potential $U(\sigma)$. After taking the pivot scale as $N_*=50$, we have a total of $5$ free parameters $m,U_0,\gamma_1,\gamma_3$ and $N_c$. By using the same trick as in \refsec{Plateau from a saddle/inflection point}, we can fix the overall energy scale to match $P_\zeta$, and switch the parameters to $r,\gamma_0,\gamma_3,N_c$. From \refeq{sip-ns1} to \refeq{sip-gnl1} and \refeq{sip-dc2}, we can see both the cosmological observables and the solution $x_*(N_*)$ are independent of $\gamma_3$ when $\gamma_3$ is large. Therefore, here we take a large $\gamma_3$ and further reduce the parameter space to $r,\gamma_0,N_c$.

\figa{\figi{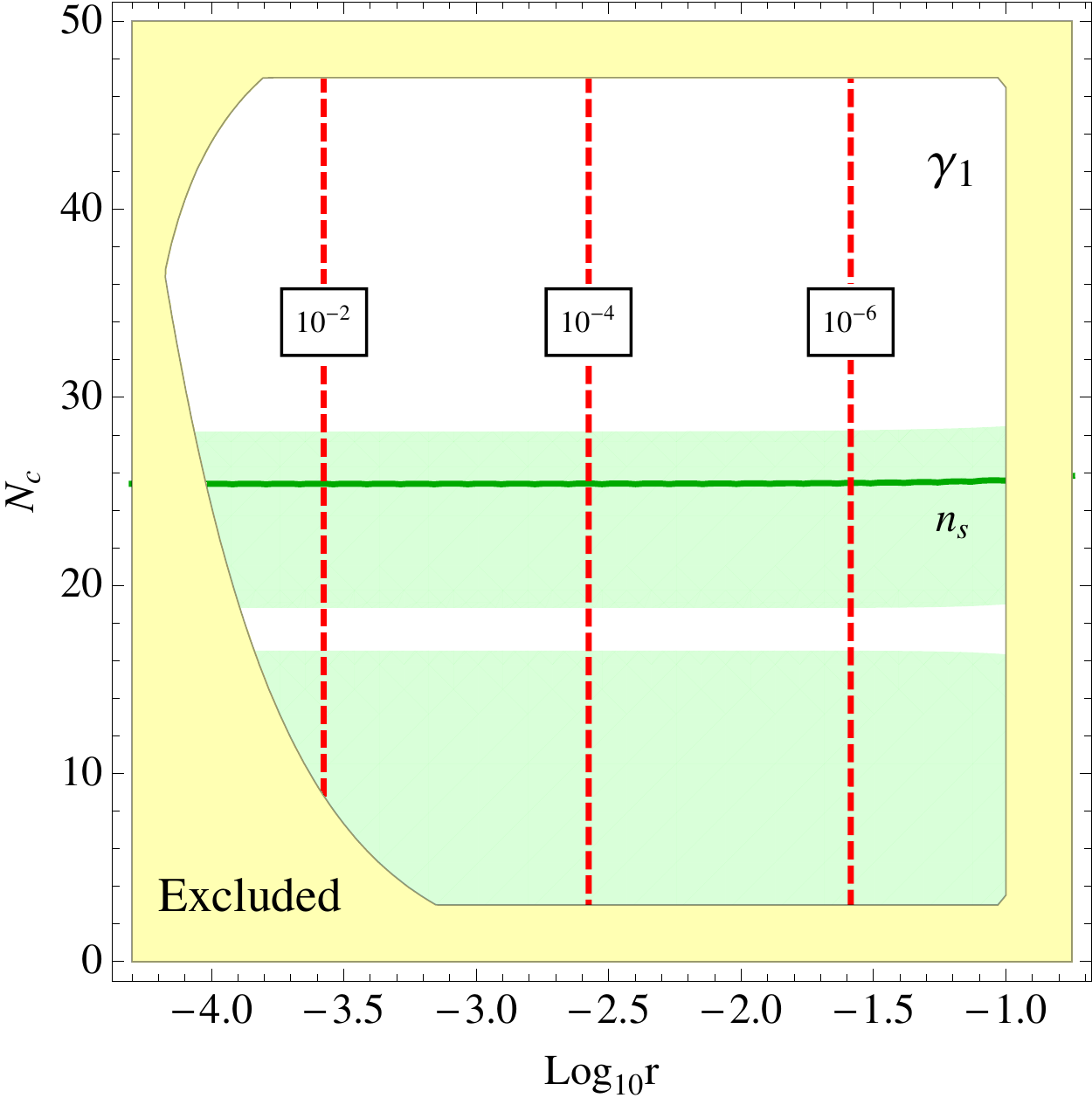}{The parameter $\gamma_1$ as a function of $r$ and $N_c$ in the new parameter space within the observational bound of $n_s$.}{0.49}\hspace{0.12in}\figi{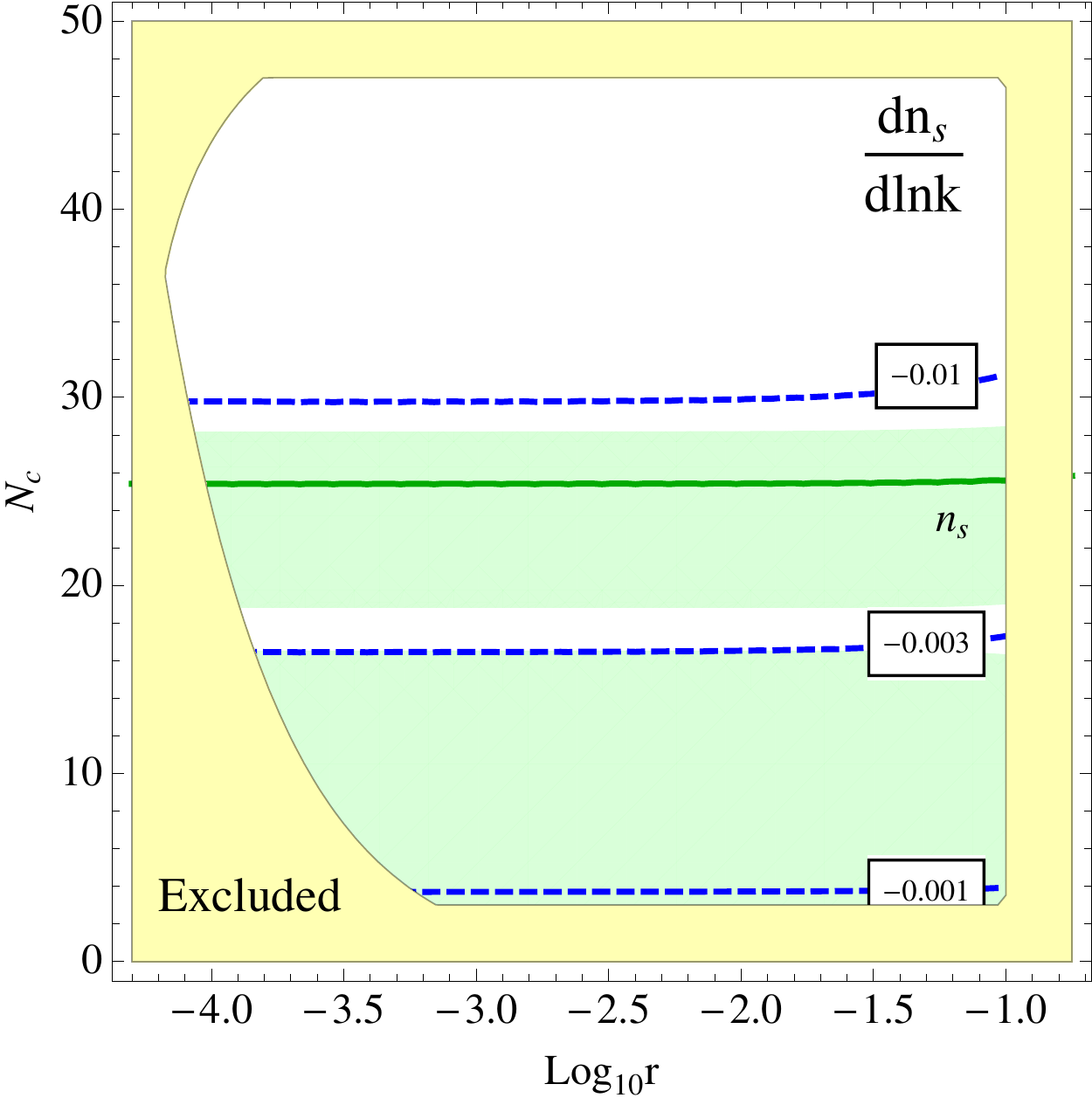}{Running of the spectral index $\dd n_s/\dd\ln k$ within the observational bound of $n_s$.}{0.49}\\
\figi{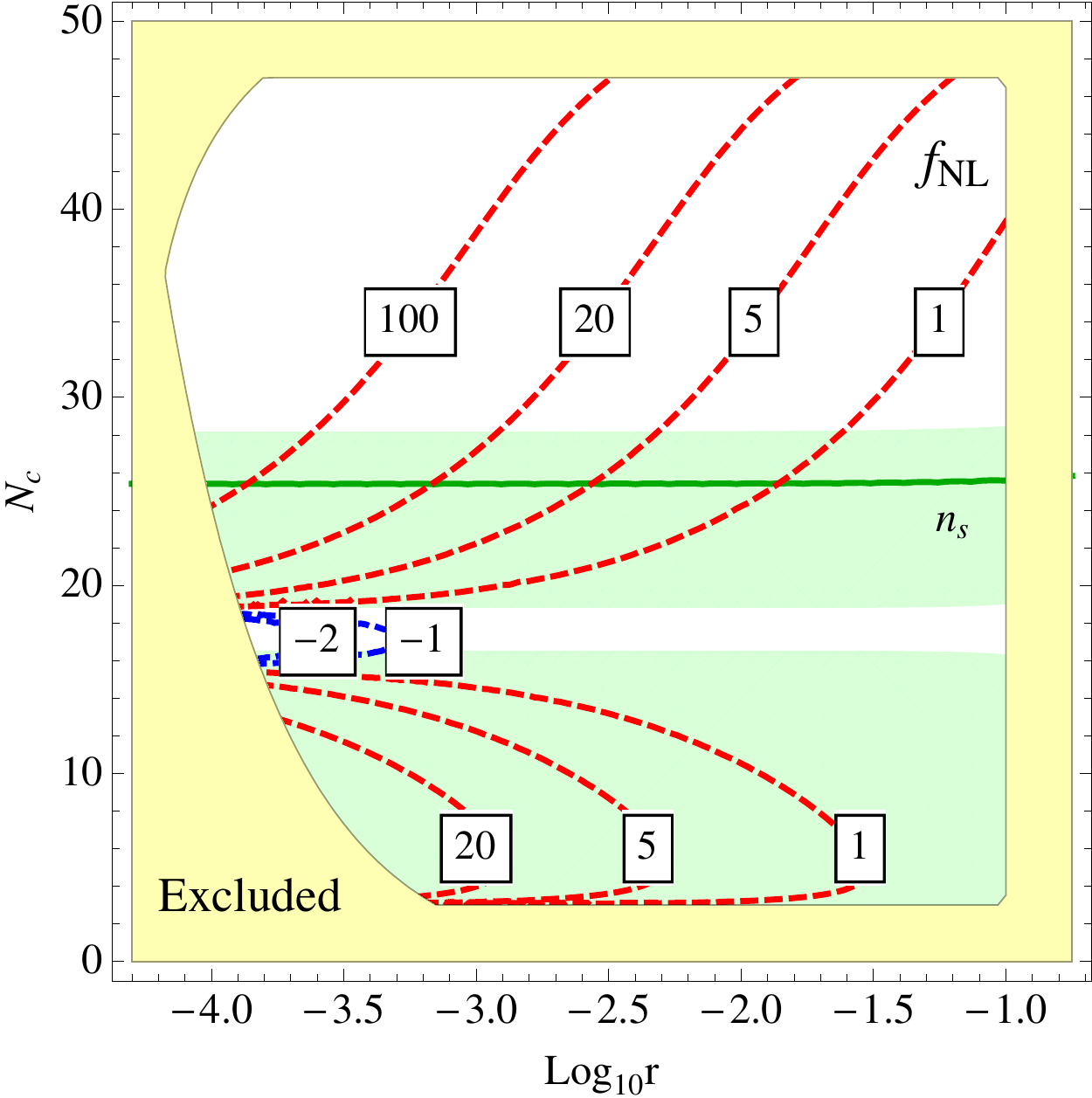}{The local bispectrum $\fNL$ within the observational bound of $n_s$.}{0.49}\hspace{0.12in}\figi{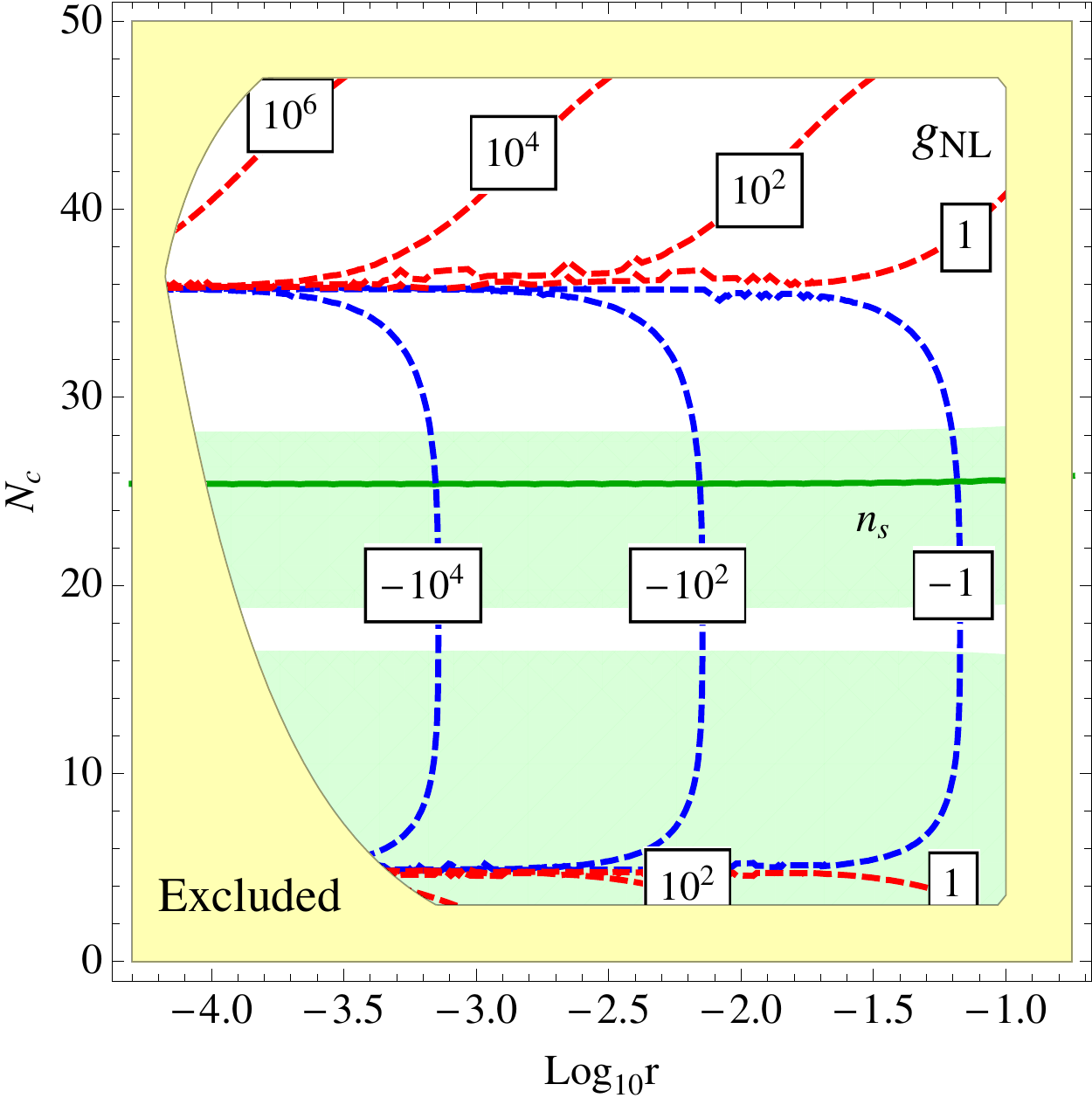}{The local trispectrum $\gNL$ within the observational bound of $n_s$.}{0.49}
}{Cosmological observables and $\gamma_1$ are drawn for the inflection point spectator field with a quadratic inflaton potential. The $x$ axis is the logarithmic of $r$, the energy density ratio of the spectator field w.r.t the total at the phase boundary. The $y$ axis is the number of e-folds of inflation from the phase boundary ``$c$'' to the end of inflation. The shaded yellow regions are excluded by the five constraints discussed in \refsec{Plateau from a step function}. The shaded green regions are observationally favored by spectral index within $n_s=0.9608\pm0.0160$, whereas the solid green lines indicate the central value. \cite{Bennett:2012fp}\label{fig-inf}}{h}
The background evolution of $\fin$ can be worked out as a function of $N$
\eq{\fin(N)=\mpl\sqrt\frac{2N+1}{4\pi}.\label{eq-sipq-phi}}
This reduces \refeq{sip-dc2} to
\eq{\arctan\frac{x\ssh}{\gamma_0}=-\arctan\frac{1}{\gamma_0}+\frac{\gamma_0(2N\sst+1)}{4}\ln\frac{2N\ssh+1}{2N\sst+1}.\label{eq-sipq-xsol}}

In \refeq{sipq-xsol} when $N_*=50$ and $\gamma_0$ are fixed, the relative displacement $x_*$ is maximized when the last term in \refeq{sipq-xsol} reaches maximum at $N_c\approx18$. Since the field $\fcu$ contributes to the spectral index by $2x_*$ as in \refeq{sip-ns1}, the spectral index would also reach maximum at $N_c\approx18$. The local bispectrum has a $-5x_*/6r$ contribution, so it should reach minimum at $N_c\approx18$. For the $m^2\phi^2$ inflaton potential we have $\xi_\phi=0$, so $\gNL$ would be positive for $x_*^2>\gamma_0^2/3$ and negative for $x_*^2<\gamma_0^2/3$.

To verify our analytical analysis and check the observational bound for the parameter space, we take the specific case of $N_*=50$, $\gamma_3=10^{10}$, $\gamma_0=0.15$ and $w=1/3$. As the power spectrum can be determined by the overall energy scale, we plot the rest of the CMB observables w.r.t the $r$ and $N_c$ coordinates in \refig{inf}. The parameter $\gamma_1$ is also shown in the transformed parameter space. The parameter space has the same exclusion conditions as discussed in \refsec{Plateau from a step function} which are shaded yellow. Here the energy scale for the spectator field, $U_0$, varies a lot from $10^{-26}\mpl^4$ to $10^{-17}\mpl^4$.

In \refig{inf}, we find agreement with the analytical analysis. The spectral index drops out of the ``$2\sigma$'' confidence level around $N_c\approx18$ because its maximum value exceeds the observed central value. Also at $N_c\approx18$, the local bispectrum reaches its minimum in the same time. The running of spectral index is typically small. The local bispectrum and trispectrum can both attain large or small values with plenty of parameter space within the current observational bound\ \cite{Bennett:2012fp,Smidt:2010ra}. In this case, the significant $\fNL$ and $\gNL$ come from the $\eta_{\sigma*}/r$ term in \refeq{c2-Nppr1}.

\secs{Conclusion}
We have derived analytically the influence of a generic spectator field $\sigma$ on the temperature anisotropy of the CMB. We evaluated the relevant cosmological perturbations seeded by the spectator field by $\delta N$ formalism, and obtained possible non-Gaussian signatures. Significant local bispectrum and trispectrum on the temperature anisotropy of the CMB can arise from the possible curvature of the spectator potential and the dilution process of $\fcu$'s decay products. Since the spectator field decays well before inflation comes to an end, the decay products of the spectator field are completely diluted away during inflation, and will not generate any isocurvature perturbations. In this respect the spectator field might come from the visible sector or from the hidden sector, and need not have any couplings to the Standard Model fields\footnote{ If the spectator oscillates or its decay products become nonrelativistic before the matter-radiation equivalence, we need sufficient e-folds of inflation after the spectator ends slow roll, i.e.\ $N_c$, to prevent any observable isocurvature perturbations. In the extreme case where the spectator decays directly into nonrelativistic dark matter, we typically require $N_c>N_*/3\sim20$ for $N_*=60$.}. 

We have considered two examples for the spectator potential where it can match the current observations.  In the first case we have considered a step potential for $\fcu$ and for inflaton we have assumed a quadratic potential. In this case $\fNL\lesssim 20$ and 
$\gNL$ is very small. The model predicts the spectral index to be around $0.98$ with a negligible running. In the second example we have considered a simple potential for $\fcu$ with an inflection point. The $\fNL$ and $\gNL$ can be as large as the upper bound allowed by the current limit~\cite{Bennett:2012fp}, while the spectral index and its running are both well within the observational bound.

To summarize, we have found a new mechanism where the spectator field can indeed provide dominant curvature perturbations, with or without a deviation from Gaussian fluctuations. This suggests that fields that decay during inflation can still play an important role, and hence should not be neglected in many plausible scenarios. For any beyond-the-Standard-Model physics that introduces plethora of hidden sectors, even though they might be dormant and decay during inflation, given sufficient conditions laid down in this paper, they can still leave their imprints in the sky.

\acknowledgments{}
We would like to acknowledge the role of the UK Particle Cosmology workshop in the dissemination of this work. AM is supported by the STFC grant ST/J000418/1.
\appendix{}
\secs{Notations}
A list of notations which are present in the final expressions can be found in \reftab{Notations}. 
\tab{Notations}{h}{|c|c|c|}{
	\hline
	Symbol&Definition&Expression\\
	\hline\hline
	$\ssh$&Value at the Hubble exit&\\
	$\sst$&Value at the phase boundary&\\
	$\sse$&Value at the end of inflation&\\
	$f$&Phase boundary condition&$\mpl^2\Vc''+8\pi(\Vc+\Vi)$\\
	$w$&Equation of state during phase II&\\
	$r$&Energy density ratio&$\Vc\sst/(\Vc\sst+\Vi\sst)\ll1$\\
	$s$&&$(\Vc\sst+\Vi\sst)/(\Vc\ssh+\Vi\ssh)<1$\\
	$s_{\fcu}$&&$\Vc\sst/\Vc\ssh<1$\\
	\hline\multicolumn{3}{|c|}{Slow roll parameters (similarly for $\fin$)}\\\hline
	$\epsilon_{\fcu}$&First order&$\mpl^2\Vc'{}^2/16\pi(\Vc+\Vi)^2$\\
	$\eta_{\fcu}$&Second order&$\mpl^2\Vc''/8\pi(\Vc+\Vi)$\\
	$\xi_{\fcu}$&Third order&$\mpl^4\Vc'\Vc'''/(8\pi(\Vc+\Vi))^2$\\
	$\lambda_{\fcu}$&Fourth order&$\mpl^6\Vc'{}^2\Vc''''/(8\pi(\Vc+\Vi))^3$\\
	$\chi_{\fcu}$&Fifth order&$\mpl^8\Vc'{}^3\Vc'''''/(8\pi(\Vc+\Vi))^4$\\
	\hline
}

\section{The expression of $\alpha$\label{sec-The alpha}}
From the definition of $\alpha$, \refeq{c12-alpha}, we calculate the integral by changing the variable to $n(\fin\sst,\fin)$, or just simplified as $n$, through the slow roll relation
\eq{\dd\fin=-\frac{\mpl^2\Vi'}{8\pi\Vi}\dd n.}
Then we get
\eq{\int_{\fin\sse}^{\fin\sst}\frac{e^{-3(1+w)n(\fin\sst,\fin)}}{\Vi'}\dd\fin=\int^{n(\fin\sst,\fin\sse)}_0\frac{\mpl^2}{8\pi\Vi}e^{-3(1+w)n}\dd n.\label{eq-c2a-int1}}
The term $1/\Vi$ in \refeq{c2a-int1} can be regarded as a function of $n$ and Taylor expanded at the phase boundary point, giving
\eq{\frac{1}{\Vi}=\frac{1}{\Vi\sst}-\frac{n}{\Vi_c^2}\left.\frac{\dd\Vi}{\dd n}\right|_{\st}+\frac{n^2}{\Vi_c^3}\left(\left.\frac{\dd\Vi}{\dd n}\right|_{\st}^2-2\left.\frac{\dd^2\Vi}{\dd n^2}\right|_{\st}\Vi\sst\right)+\cdots.\label{eq-c2a-Vi}}

For calculations of $n$-point correlation functions, it is safe to truncate the series expansion up to $(n-1)$-th order. We assume the phase boundary $\st$ and the end of inflation $\se$ are well separated by several e-folds so $n(\fin\sst,\fin\sse)\gtrsim O(3)$. In this case we can safely extend the upper bound of the integral to infinity because of the exponential suppression in $n$. Then the value of $\alpha$ can be worked out order by order
\eq{\alpha=\alpha\sups{(1)}+\alpha\sups{(2)}+\alpha\sups{(3)}+\cdots.\label{eq-c2a-a0}}

For the local bispectrum, we need the first two terms in \refeq{c2a-a0}, which are
\eq{\alpha\sups{(1)}=-\frac{2\epsilon_{\fin}\sst}{3(1+w)(1-r)^2},}
and
\eq{\alpha\sups{(2)}=-\frac{4\epsilon_{\fin}\sst}{9(1+w)^2(1-r)^3}\left(\frac{3\epsilon_{\fin}\sst}{1-r}-\eta_{\fin}\sst\right),}
where $\eta_{\fin}\equiv\mpl^2\Vi''/8\pi(\Vc+\Vi)$ is the second order slow roll parameter for $\fin$ when both fields present.

For the trispectrum calculation, $\alpha$ needs to be calculated up to third order. Similar derivations yield the third order component for $\alpha$
\eq{\alpha\sups{(3)}=-\frac{4\epsilon_{\fin}\sst}{27(1+w)^3(1-r)^4}\left(\frac{10\epsilon_{\fin}\sst}{1-r}\biggl(\frac{3\epsilon_{\fin}\sst}{1-r}-2\eta_{\fin}\sst\biggr)+2\eta_{\fin c}^2+\xi_{\fin}\sst\right),}
and here
\eq{\xi_{\fin}\equiv\frac{\mpl^4\Vi'\Vi'''}{64\pi^2(\Vc+\Vi)^2}\label{eq-c2a-xifin}}
is the third order slow roll parameter for $\fin$.

\bibliographystyle{jcap}
\bibliography{Main}
\end{document}